# Single 5-nm quantum dot detection via microtoroid optical resonator photothermal microscopy

Shuang Hao[1], Sartanee Suebka[1], and Judith Su[1,2,*]
[1]Wyant College of Optical Sciences, The University of Arizona
[2]Department of Biomedical Engineering, The University of Arizona
*judy@optics.arizona.edu

**Label-free detection techniques for single particles and molecules play an important role in basic science, disease diagnostics, and nanomaterial investigations. While traditional fluorescence-based methods offer powerful tools for single molecule detection and imaging, they are limited by a narrow range of molecular probes and issues such as photoblinking and photobleaching. Photothermal microscopy has emerged as a label-free imaging technique capable of detecting individual nanoabsorbers with high sensitivity. Whispering gallery mode microresonators can confine light in a small volume for enhanced light-matter interaction and thus are a promising ultra-sensitive photothermal microscopy platform. Previously microtoroid optical resonators were combined with photothermal microscopy to detect 250 nm long gold nanorods. Here, we combine whispering gallery mode microtoroid optical resonators with photothermal microscopy to spatially detect 5 nm diameter quantum dots (QDs) with a signal-to-noise ratio (SNR) exceeding $10^4$. To achieve this, we integrated our microtoroid based photothermal microscopy setup with a low amplitude modulated pump laser and utilized the proportional-integral-derivative (PID) controller output as the photothermal signal source to reduce noise and enhance signal stability. The measured heat dissipation of these 5 nm QDs is below the detectable level from single dye molecules, showcasing the high sensitivity and discrimination capabilities of this platform. We anticipate that our work will have application in a wide variety of fields, including the biological sciences, nanotechnology, materials science, chemistry, and medicine.**

The detection of individual particles and molecules has had impact in understanding protein dynamics[1], DNA[2] and RNA[3,4] analysis, cellular imaging, nanotechnology and nanomaterials, and biomedical diagnostics, among other fields. Traditional single-molecule fluorescence-based detection methods such as Stochastic Optical Reconstruction Microscopy (STORM) or Photo-activated localization microscopy (PALM) are powerful tools to study molecular processes. Such techniques are widely used and valued for their low background noise and high sensitivity[5–7]. Fluorescence techniques, however, are limited as they are restricted to a narrow range of molecular probes with high fluorescence quantum yields. Additionally, issues such as photoblinking[8–10] and photobleaching[9] limit their effectiveness.

As such, photothermal microscopy has emerged as a label-free non-invasive imaging technique. Photothermal microscopy measures localized variations in the refractive index of a sample's surroundings. These variations result from the absorption of light by sample components, which in turn induce temperature changes in the surrounding region[11]. Photothermal microscopy can provide insight into the optical and thermal properties of materials[12–15] and biological structures[16–18]. Its high sensitivity renders it advantageous across a diverse array of applications, including material science[11,19], nanotechnology[20–22], biological imaging[23,24], and thermal metrology[25–27].

Currently, photothermal heterodyne imaging (PHI)[28] is the predominant method for photothermal microscopy of single nano-objects. PHI can detect gold nanoparticles as small as 1.4 nm in diameter[28] with a signal-to-noise ratio (SNR) over 10 and measure the absorption spectra of single ~ 7 nm quantum dots (QDs) at room temperature[29] with a SNR < 10. By using multiple pump beams, PHI can perform multiplexed imaging, enabling simultaneous targeting and detection of gold and silver particles[30] or dynamic imaging of mitochondria and lysosomes in living cells[17]. Nanomechanical silicon nitride drums have also been used as mechanical transducers for photothermal imaging of single dye molecules, specifically Atto 633, with a heat dissipation of 6.3 pW[31]. Although photothermal microscopy coupled with mechanical transducers exhibits high sensitivity, it is limited to operating under high-vacuum conditions.

Here, to perform ultra-sensitive photothermal imaging in ambient air at room temperature, we use whispering gallery mode (WGM) microtoroid resonators as detectors in photothermal microscopy and achieve single 5 nm QDs detection with an SNR over $10^4$ (see supplemental information for calculation) and with a simpler system and alignment requirement than PHI. WGM microtoroid optical resonators can measure small temperature changes induced from the heat dissipation of molecules. They are a class of optical microcavities known for their ultra-high quality (Q) factors, making them suited for a diverse set of applications[32], including single molecule detection[33–35], biochemical detection,[36,36–42] and frequency comb generation[43–46]. The Q-factor characterizes the efficiency and energy loss within the resonator and is mathematically expressed as $Q = \lambda_0/\Delta\lambda$, where $\lambda_0$ is the WGM resonant wavelength and $\Delta\lambda$ is the full width at half maximum linewidth (FWHM) of the WGM. The ultra-high Q factor of WGM resonators results in a small mode volume[47–49], which in turn greatly enhances light-matter interaction. In addition, the narrow resonances of WGM enable precise measurement of the resonance shift. These high Q factors can enable photothermal microscopy with high sensitivity and precision.

Among various types of WGM optical resonators, such as microspheres[50–52], microdisks[53–55], microbubbles[56] and microtoroids[57–59], microtoroids stand out due to Q-factors[60] in excess of $10^8$, and a flat disk plate that facilitates the placement of nanoabsorbers. Previously, photothermal spectroscopy of single gold nanorods (250 nm × 25 nm) was demonstrated using microtoroids[27]; however, detecting smaller particles has proven challenging due to the high background signal generated from the microtoroid's silicon pillar. To overcome this limitation, all-glass microtoroids[61] have been designed and fabricated to minimize unwanted absorption from the supporting pillar. The all-glass fabrication process, however, leads to a sacrifice in the Q-factor, resulting in values around $10^6$.

Alternatively, we have opted for an easy fabrication method for microtoroids to both expand the effective detection area and preserve the high Q-factor. Here, we demonstrate photothermal microscopy based on a re-etched microtoroid and low amplitude modulation (AM) frequency of the pump laser. We use the proportional-integral-derivative (PID) controller output signal to measure the resonance shift instead of the error signal as was previously used[27]. The re-etching process of the microtoroid significantly decreases the size of the microtoroid pillar, providing two benefits: (1) it reduces the area with high background signal and (2) provides better thermal isolation of the microtoroid, enhancing the photothermal effect. In our approach, nanoparticles are deposited onto the top surface of the microtoroid. Upon illumination with a pump light at a wavelength of 405 nm, absorption of light by the nanoparticles leads to localized heating and dissipation within the resonator. In the case of the fused silica microtoroid, it possesses a positive thermal expansion coefficient $(5.5 \times 10^{-7}\ K^{-1})$[62] and a positive thermo-optic coefficient $(8.6 \times 10^{-6}\ K^{-1})$[63]. Both positive coefficients contribute to the optical path length increasing, consequently leading to a redshift in resonance wavelength. The photothermal signal is directly related to the heat dissipation from the optical absorbing particles. The shift in the WGM resonance is proportional to the absorption cross-section of the nanoparticles. To detect these resonance shifts, we utilize an additional probe laser with a wavelength of 780 nm, which is intentionally detuned far away from the pump laser to avoid interference from nanoparticle or molecule absorption. Compared to the high frequency amplitude modulation of the pump laser used by others, we use low frequency modulation of the pump laser to enable the frequency lock-in system to track the oscillating resonance shift signal closely and accurately, improving the SNR of the photothermal signal.

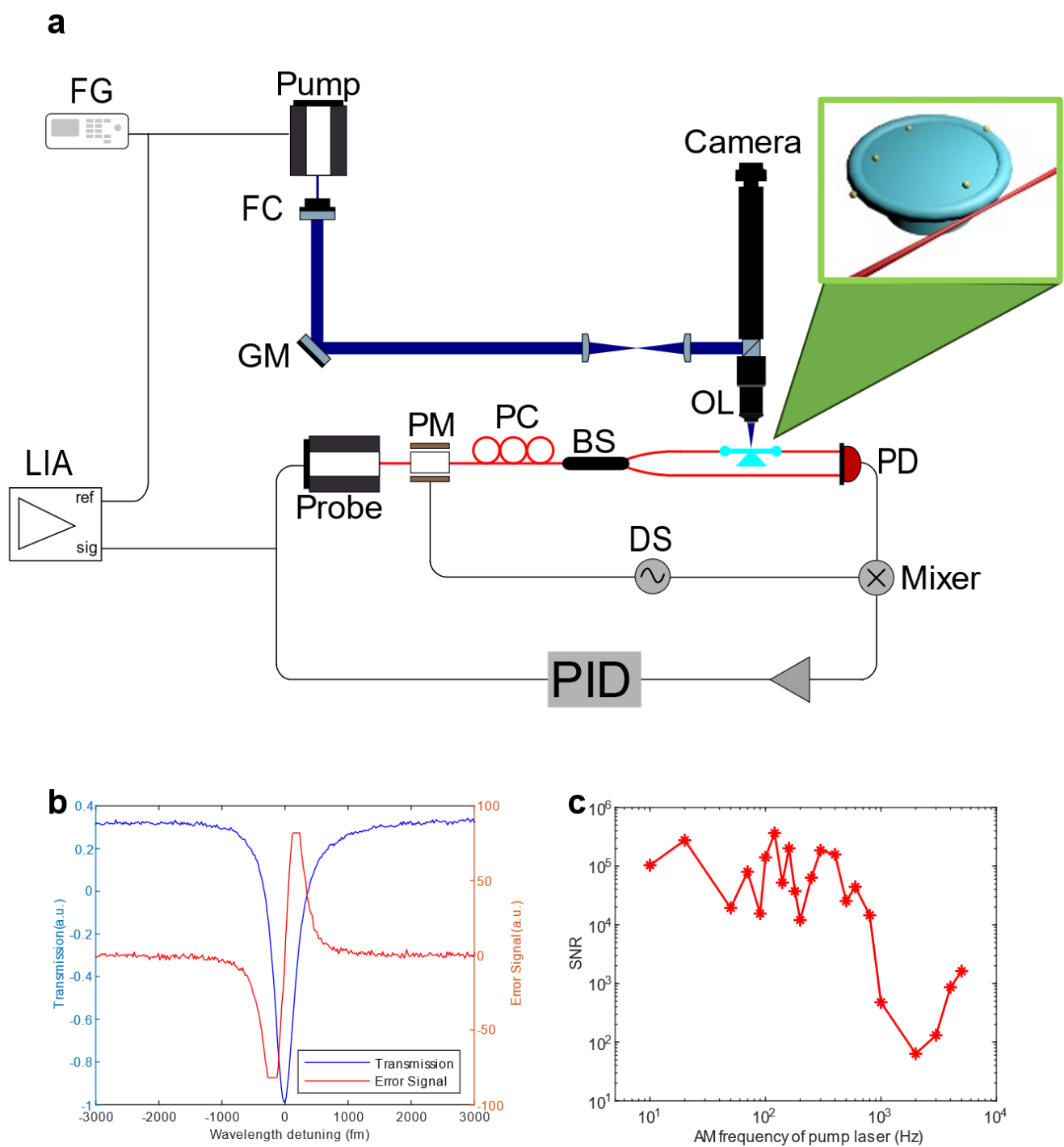

**Figure 1 Photothermal microscopy based on microtoroid. a**, Photothermal microscopy system setup. The top right is the particle placed microtoroid coupling to tapered fiber. Key components include FG (Function Generator), FC (Fiber Collimator), GM (Galvo Mirror), LIA (Lock-in Amplifier), PM (Phase Modulator), PC (Polarization Controller), BS (fiber Beam Splitter), OL (Objective Lens), PD (Photodetector), DS (Dither Signal), PID (Proportional-Integral-Derivative controller), Probe laser and Pump laser. **b**, The resonance transmission of the microtoroid resonance (blue curve) is acquired by the probe laser scanning. The Q factor of the resonance is $1.86 \times 10^6$. The error signal (red curve) equals to zero at the resonance peak wavelength. **c**, The SNR of the photothermal signal of DiagNano (DN) 800 QDs (size:5~6nm). The time constant is set as 1 s in lock-in amplifier.

In previous work, we developed a system called Frequency Locked Optical Whispering Evanescent Resonator (FLOWER)[64–68], which combines optical microcavities with frequency locking and data processing to enable the detection of single macromolecules. Unlike conventional methods that involve scanning the wavelength of a tunable probe laser, which can be limited by scanning speed and hinder real-time tracking of resonance shifts, FLOWER uses a Pound-Drever-Hall (PDH) like technique[69] to reduce the response time and enhance the accuracy of resonance shift measurement.

Our experimental setup integrates FLOWER with photothermal microscopy. The experimental setup shown in Fig. 1a involves coupling a probe laser with a wavelength of 780 nm into the microtoroid resonator through a tapered optical fiber. Free-space coupling of light into the toroid is also possible.[67,70] To optimize the coupling efficiency between the tapered fiber and microtoroid, we control the polarization of the probe laser using a polarization controller (PC). A 25 MHz oscillation dither signal (DS) is applied to drive the phase modulator, resulting in phase modulation of the probe laser. Subsequently, the phase-modulated probe laser is split into two arms through a fiber-coupled beam splitter (BS): one arm carries the signal light coupled into the microtoroid, while the other serves as the reference light. Both the signal and reference light are then received by a high-bandwidth balanced photodetector (PD). This balanced detector plays a crucial role in removing common noise from the probe laser. By multiplying the balanced photodiode's electrical output signal by the dither signal and time-averaging, we obtain an error signal that is proportional to the wavelength detuning between the probe laser and the WGM resonance of the microtoroid. Fig. 1b shows the resonance transmission spectra and its corresponding error signal, where the error signal becomes zero at the resonance when the probe laser perfectly matches the microcavity WGM resonance. Utilizing this error signal in a feedback loop with the tunable laser controller, the PID controller minimizes the error signal, thereby ensuring accurate and stable frequency locking of the probe laser to the WGM resonance.

In the PID control system, the output of the PID controller represents the resonance shift signal. Conversely, the error signal reflects the wavelength detuning between the probe laser and the WGM resonance. The oscillatory resonance shift with low frequency can be tracked more effectively by the PID controller output signal instead of the error signal. The PID method boasts an SNR exceeding tenfold that of the error signal method (Supplementary Information Section 1). During photothermal imaging experiments involving nanoparticles, we deposit QDs or Au nanospheres onto the microtoroid. However, the introduction of these nanoparticles induces additional losses, leading to a reduction in the Q-factor of the microtoroid. In Fig. 2b, the Q-factor of the microtoroid with Au nanospheres is measured to be $1.86 \times 10^6$. The selection of this specific Q-factor resonance is guided by careful consideration of its impact on the photothermal signal detection range. While higher Q factors generally lead to increased sensitivity, they may limit the effective tracking range of the resonance shift and risk destabilizing the frequency locking due to high photothermal signals. The Q-factor values achieved in the $10^6$ range are comparable to those observed in single-molecule detection using WGM resonators[71–73]. Choosing a resonance with a Q factor in this range strikes a balance, enabling imaging of the entire microtoroid without saturation distortion while preserving good sensitivity. In future experiments aimed at detecting small single molecules with low absorption, selecting a resonance with a higher Q factor can further enhance sensitivity.

In our photothermal microscopy system (Fig. 1a), an additional continuous wave (CW) laser emitting at 405 nm serves as the pump laser. This specific wavelength resides at the boundary between visible and UV light spectra, aligning with the characteristic absorptions exhibited by a wide range of molecules and particles. This wavelength choice guarantees the generation of a robust photothermal signal. Significantly, the pump laser's wavelength is detuned from that of the probe laser, preventing any potential interference between the two.

The photothermal microscopy involves the transmission of two beams of light into the microtoroid resonator. The first beam is the probe laser, which couples to the microtoroid via a tapered fiber. The second beam is the pump laser, propagating in free space which illuminates the microtoroid through a lens system. The pump laser output is directed through a fiber collimator (FC) to convert it into a free-space collimated beam. This collimated beam is then transmitted through a galvo mirror (GM) scan system and focused on the top surface of the microtoroid using a 60X objective lens. The relay lens, which is comprised of a scan lens and a tube lens, is positioned between the GM and the cubic beam splitter. The GM scan system allows for the variation of the incident angle, resulting in changes in the laser spot position on the microtoroid plane. By employing a two-axis rotation for the GM, a 2D spatial scan of the pump laser on the microtoroid can be achieved. The pump laser is amplitude modulated with a 203.7 Hz oscillating signal from a function generator. The resulting 203.7 Hz AM signal in the resonance shift is detected by FLOWER. The amplitude of this oscillation directly correlates with the heat dissipation caused by the pump laser beam, effectively representing the photothermal signal. Fig. 1c depicts the SNR of the photothermal signal of single 5-6nm QD as a function of the AM frequency. Remarkably, the SNR remains above $10^4$ when the AM frequency is below 200 Hz. To capitalize on the high SNR and achieve accurate real-time tracking of the photothermal signal, we utilize a lock-in amplifier operating at the AM frequency of 203.7 Hz (Additional details regarding the AM frequency selection are provided in Supplementary information Section 1). This combination of techniques significantly improves the SNR of the image acquired by the photothermal microscopy based on the re-etched microtoroid optical resonator.

## Results
### Imaging of Au nanospheres
The use of gold nanoparticles in conjunction with antibody labeling is a valuable technique for biomolecule detection[74,75]. Gold nanoparticles can serve as markers or tags that can be easily visualized and detected due to their distinct optical properties. Therefore, a 100 nm Au nanosphere was first selected as a target particle. The intensity of the 405 nm pump laser spot is $17.4\ KW/cm^2$, which is significantly below the melting point of the Au particles. A 2D scan of the microtoroid is first performed using the pump laser beam. The amplitude of the resonance shift is recorded with a double-locking mechanism, both from FLOWER and from the lock-in amplifier. The photothermal signal of each pixel during the 2D scan is the same and is longer than the time constant of the lock-in amplifier. The Au nanospheres are placed on the microtoroid using an aerosol generator. The 100 nm Au nanosphere physically bind on the microtoroid. Fig. 2a presents the photothermal map of the entire microtoroid with Au nanospheres. The microtoroid structure consists of glass material for the disk and toroidal rim parts, which exhibit minimal absorption of the 405 nm pump laser. The supporting pillar of the microtoroid is made of silicon, which has high absorption. This configuration results in a low (similar to background) photothermal signal being generated by the microtoroid itself. In contrast, the pillar produces a strong and distinct photothermal signal. As a result, the effective detection area for particles is primarily the microtoroid area excluding the pillar. To expand the effective area for nanoparticle detection, secondary etching is employed during microtoroid fabrication, reducing the pillar size. Additionally, the smaller pillar contributes to the heat insulation of the microtoroid, thus improving the sensitivity of the photothermal signal. In Fig .2a, multiple photothermal hotspots are distributed around the rim of the microtoroid, representing the presence of 100 nm Au nanospheres. The absorption cross section of 100 nm Au nanosphere is calculated as $\sigma_{Au} = 1.9739 \times 10^4 \, nm^2$. The heat dissipation is $P_{heat} = 3.4277 \times 10^6 pW$. The calculation details are provided in Supplementary information Section 2. A fine photothermal map of an individual 100 nm Au nanosphere is shown in Fig 2b. This result highlights the ease with which the 100 nm Au nanosphere can be detected using photothermal microscopy. To further explore the detection limits of this photothermal microscopy system, smaller nanoparticles were then chosen as target particles.

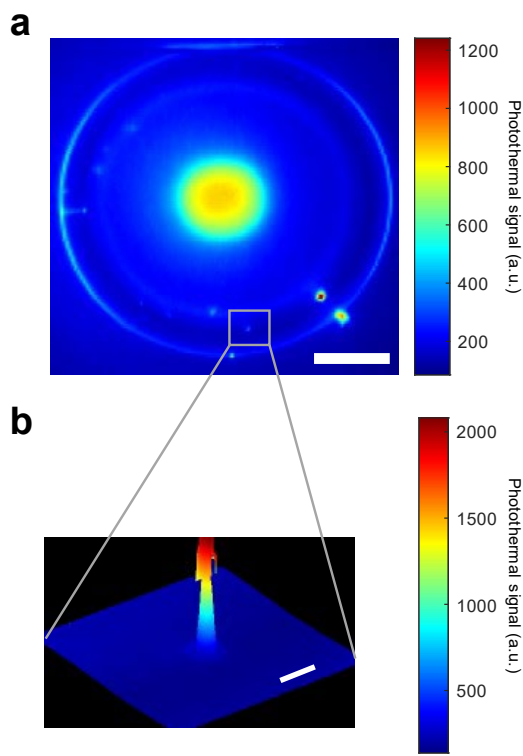

**Figure 2. Photothermal map of the microtoroid with Au nanosphere binding. a**, Coarse photothermal map of the whole microtoroid. The diameter of Au nanosphere on the microtoroid is 100 nm. Scale bar, 20 μm. **b**, Fine photothermal map of single 100 nm Au nanospheres marked in **a**. Scale bar, 2 μm.

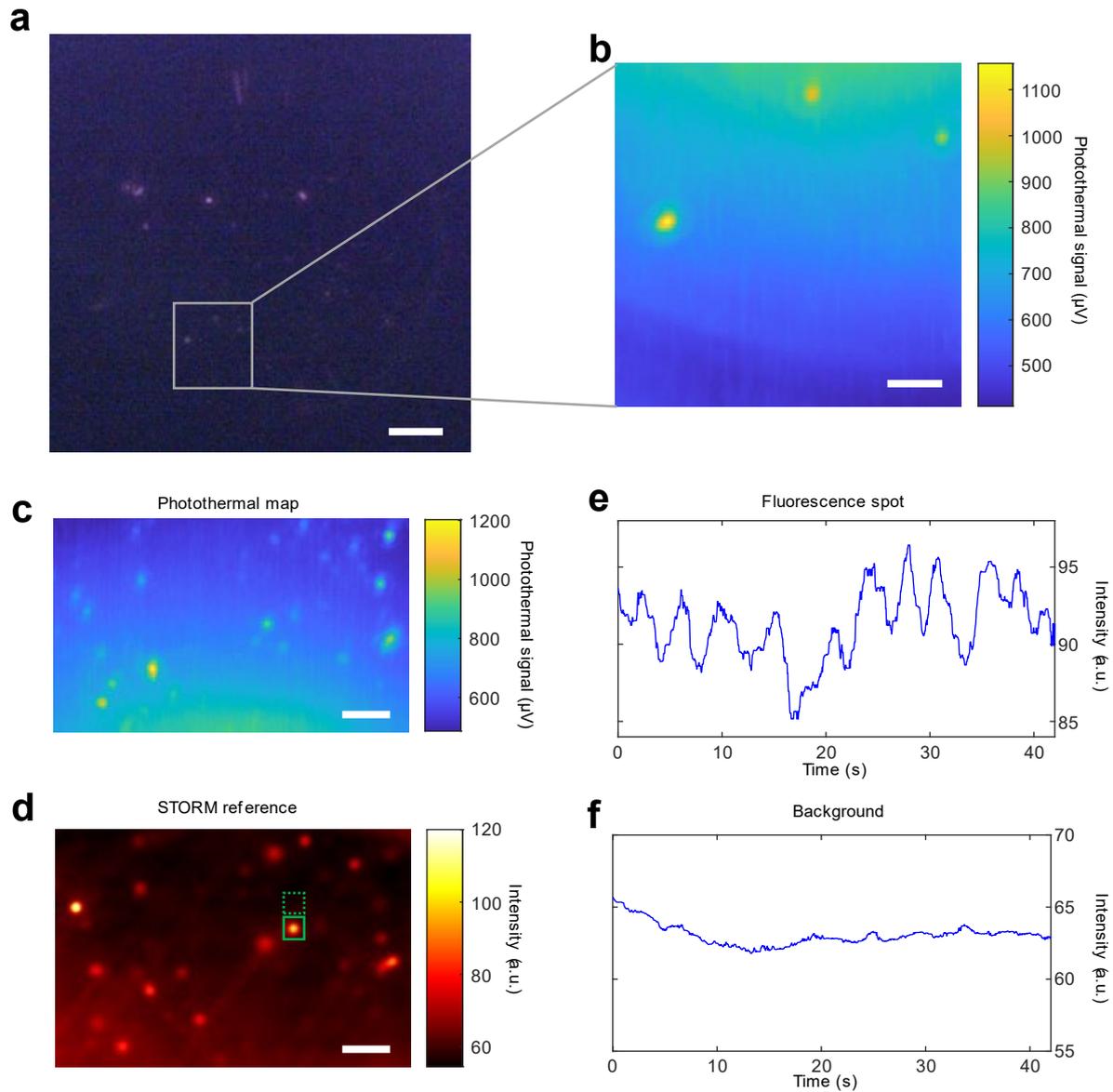

**Figure 3**. **Photothermal image of Qdot 800 QDs (size:18~20nm). a**, Fluorescence image of Qdot 800 QDs on the microtoroid. Scale bar, 10 µm. **b**, Photothermal image of three individual Qdot 800 QDs in the gray marked area in **a**. Scale bar, 2 µm. **c**, Fine photothermal map of the Qdot 800 QDs on another microtoroid. Scale bar, 3 µm. **d**, The superimposed frames image illustrating the same area as shown in **c**, acquired through STORM imaging. Scale bar:3 µm. **e**, Intensity profile of the spot within the solid green square region in **d**. **f**, The background signal captured within the green dashed square region of **d**.

## Imaging of QDs
To demonstrate the high sensitivity of photothermal microscopy for single nanoparticles, we specifically opted for Qdot 800 QDs (Thermo Fisher Scientific) as our target particles based on their well-established and recognized characteristics. Firstly, the QDs emit stable fluorescence light at a peak wavelength of 793 nm. The fluorescence image can serve as a reference image for the photothermal map. Secondly, unlike fluorescence dyes, QDs do not undergo photobleaching, enabling longer observation times and stable fluorescence images. Thirdly, Qdot 800 QDs exhibit strong absorption in the UV spectrum, which matches the 405 nm wavelength of the pump laser. The diameter of the Qdot 800 QDs ranges from 18 nm to 20 nm. Fig. 3a displays the fluorescence image of Qdot 800 QDs on a microtoroid, covering the entire WGM resonator and serving as a reference for subsequent photothermal imaging. In Fig. 3b, three QDs are scanned using photothermal microscopy within the marked region indicated in Fig. 3a. The resulting photothermal image reveals three hot spots in the disk area of the microtoroid, corresponding to the three fluorescence spots observed in Fig 3a. To confirm that the photothermal signals originate from single QDs, a fine photothermal map of Qdot 800 QDs on another microtoroid is presented in Fig. 3c. Additionally, the same microtoroid is imaged using STORM, capturing the blinking behavior of the QDs. The superimposed frames of the STORM video form the fluorescence image shown in Fig. 3d. The characteristic single step blinking behavior of single QDs is evident in Fig. 3e, while the background signal from the STORM fluorescence video is depicted in Fig. 3f. Both the fluorescence spot intensity and background signal are filtered by a median filter. The intensity of the single step blinking signal surpasses the background noise level, providing evidence for the presence of a single absorber and emitter. The sensitivity of the photothermal signal depends on the location of the absorber (Supplementary information Section 3). The sensitivity near the equator of the microtoroid is higher than in the location near the pillar. However, when selecting photothermal spots in the disk area while excluding the small pillar, the sensitivity does not significantly vary with changes in location. Photothermal microscopy enables precise identification of target locations on the microtoroid and accurate measurement of the absorption cross-section.

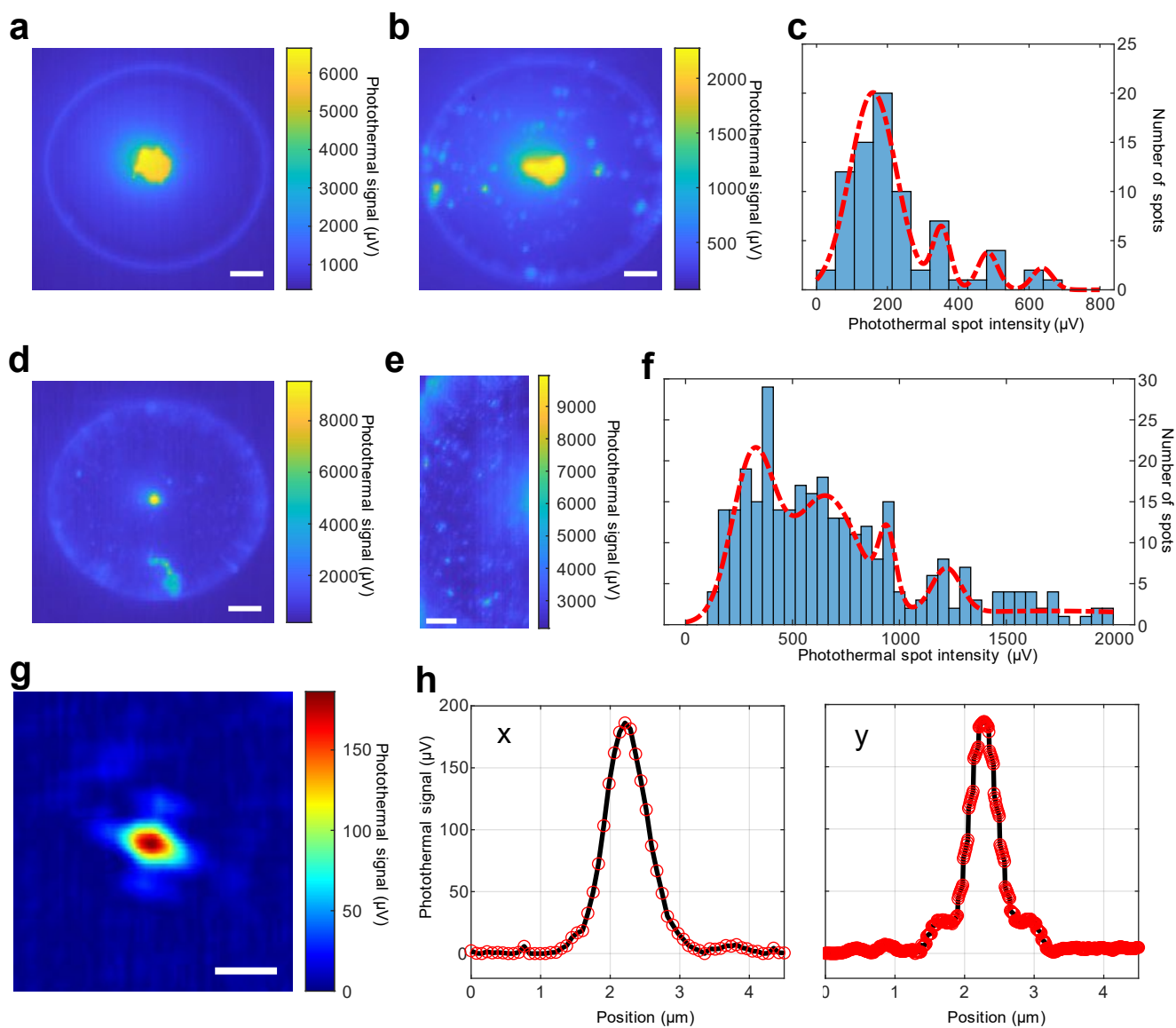

**Figure 4** Photothermal map comparison of microtoroid with QDs. **a**, Photothermal map of the microtoroid within the same chemical process, but without any QDs. Scale bar, 10 μm. **b**, Photothermal map of microtoroid with DN 800 QDs (size:5-6nm). Scale bar, 10 μm. **c**, Histogram of spot maximum intensities for DN 800 QD in **b**. **d**, Photothermal map of microtoroid with mixed QDs of DN 800 QDs and Qdot 800 QDs (size:18-20nm). Scale bar, 10 μm. **e**, Part of the microtoroid in d is photothermally imaged with high spatial resolution. Scale bar, 5 μm. **f**, Histogram of spot maximum intensities for mixed QDs in **d**. The red curves in **c** and **f** are GMM fitting. **g**, Photothermal spot of the single DN 800 QD with background signal removed. Scale bar, 1 μm. **h**, Profile cut through the photothermal peak of the single QD in both the x and y directions, as indicated in **g**.

To explore the detection limits of photothermal microscopy, we employed a smaller QDs, DiagNano (DN) 800 (CD Bioparticles). One concern was the possibility of chemical contamination on the microtoroid during the coating process, which could contribute to undesired photothermal signals. To address this, we performed a control experiment by handling the microtoroid with the same coating process but without introducing any QDs. The resulting photothermal map of the control group as shown in Fig. 4a, exhibited no detectable photothermal spots. This finding provided evidence that either there was no contamination on the microtoroid during the chemical coating process or, if present, the remaining chemical material had minimal absorption at the 405 nm wavelength, thus contributing little to the photothermal signal. Next, we conducted photothermal imaging of DN 800 QDs on the microtoroid, (Fig. 4b). Although we attempted to observe the DN 800 QDs using fluorescence microscopy or STORM, their small absorption cross-section and lower quantum yield made it challenging to observe their fluorescence or photo blinking behavior of individual QDs. Here, we analyzed the fine photothermal map of the microtoroid (Fig. 4b) and constructed a histogram of the photothermal spot intensity (Fig. 4c). Notably, individual nanoparticles exhibit a monomodal distribution in the histogram of photothermal spot[12,76], while the presence of both aggregated QDs and single QDs on the microtoroid resulted in multiple peaks. Each peak followed a Gaussian distribution, representing distinct orders of QD aggregates. The peak with the lowest mean intensity indicated individual QDs, while the peak value of higher-order aggregates was the product of the order number and the distribution peak value of an individual QD. This relationship facilitated the differentiation of different QD aggregates and individual QDs based on their photothermal spot intensity distribution. To perform the analysis, we employed a Gaussian Mixture Model (GMM) with the iterative Expectation-Maximization (EM) algorithm to fit the histogram (Fig. 4c). Our results revealed that the photothermal signal of a single DN 800 QD measured 159.6±65.3 µV, representing 80.0% of the 77 spots. Furthermore, aggregates of two QDs generated a photothermal signal of 352.7±25.0 µV (9.5% of the spots), aggregates consisting of three QDs exhibited a photothermal signal of 483.7±28.2 µV (6.6% of the spots), and four QDs exhibited a photothermal signal of 637.5±28.8 µV (3.9% of the spots). The photothermal spots primarily corresponding to individual QDs, constituted 80% of the total observed spots, while higher-order DN 800 QD aggregates showed diminishing proportions as the order number increased. Moreover, aggregates with more than four QDs were scarce on the microtoroid surface (Fig. 4b). We note that detecting such small QDs using SEM would be a challenge due to the presence of other contaminated nanoparticles that would make the QDs hard to find. In FLOWER based photothermal microscopy, such particles do not appear as they do not have a strong photothermal signal.

| | Quantum yield | Molar extinction coefficient ($M^{-1}cm^{-1}$) | Absorption cross section ($nm^2$) | Fraction of heat dissipation | Absorbed Power (pW) | Heat dissipation (pW) |
|---|---|---|---|---|---|---|
| Qdot 800 | 62% | 8.0×10⁶ | 3.0599 | 68.34% | 531.35 | 363.10 |
| DN 800 | 15.64% | 1.28×10⁶ | 0.4896 | 92.80% | 77.38 | 71.25 |

**Table 1 Optical Properties of Qdot 800 and DN 800 QDs.**
(Derivation details are provided in the Supplementary Information)

To further evaluate the discriminating capabilities of photothermal microscopy for different particles, we applied a mixture of DN 800 and Qdot 800 QDs to the microtoroid surface. The resulting photothermal image of the entire microtoroid is presented in Fig. 4d, with a finer photothermal image of the left part shown in Fig. 4e. Fig. 4e exhibits photothermal spots with varying intensities, where strong photothermal spots correspond to Qdot 800 QDs and weaker spots represent DN 800 QDs, based on their cross-section absorption characteristics detailed in Table 1. To quantitatively analyze the photothermal spots, we constructed a histogram of photothermal spot intensities in Fig. 4f and employed the same fitting methodology to identify five Gaussian distribution peaks. Specifically, photothermal spots from single Qdot 800 QDs exhibited an intensity of 1696.5±793.7 µV, constituting 21.0% of the 327 spots. In comparison, the photothermal signal from single DN 800 QDs was measured to be 316.8±99.9 µV, accounting for 31.3% of the spots. It is noteworthy that the photothermal signal of single DN 800 QDs differs between Fig. 4c and Fig. 4f, primarily due to sensitivity variations resulting from the microtoroid's pillar size. The smaller supporting pillar in Fig. 4d provides better thermal isolation, contributing to enhanced PT sensitivity. Moreover, two DN 800 QD aggregates displayed a photothermal signal of 653.2±153.3µV (36.3% of the spots), while three DN 800 QD aggregates exhibited a photothermal signal of 942.7±39.8µV (5.4% of the spots). Additionally, four DN 800 QD aggregates demonstrated a photothermal signal of 1219.3±69.0µV (6.0% of the spots). Consequently, the photothermal signal ratio of Qdot 800 QDs to DN 800 QDs was approximately 5.36. Remarkably, this experimental photothermal signal ratio for single QDs closely corresponds to the theoretically calculated heat dissipation values detailed in Table 1, exhibiting a difference of merely 5.2%.

Employing this analysis of photothermal intensity, we selected a single DN 800 QD on the microtoroid for high-resolution scanning (x direction: 75 nm/pixel, y direction: 9.375 nm/pixel). The resulting photothermal image is presented in Fig. 4g, with both x and y directions indicated in Fig.4h. Typically, after performing a high-resolution photothermal spot scan and adjusting the microtoroid position to match the pump laser's focal point, photothermal peak intensities will increase compared to the fine photothermal map of multiple spots. In Fig. 4g, the photothermal peak's value measures 186.26 µV. For a single DN 800 QD (size: 5-6 nm), this yields a SNR of $1.17 \times 10^4$. Calculations based on the experimental setup information establish the heat dissipation of a single 5 nm DN 800 QD at 77.4 pW. The high SNR observed in the context of a single 5 nm QD using our photothermal microscopy underscores our potential to detect single small molecules.

## Conclusion

In conclusion, we demonstrated that FLOWER based photothermal microscopy using re-etched microtoroids can detect single nanoparticles, as small as 5 nm QDs, with SNR exceeding 10⁴. The heat dissipation of these 5 nm QDs is 77.4 pW, which is below the 0.34 nW heat dissipation power from single dye molecules in photothermal detection[77], providing evidence of our photothermal microscopy's ability to detect single molecules. Despite the higher photothermal sensitivity of the toroid area, we opted to select the photothermal disk area for detecting QDs and measure their absorption based on the photothermal signal. This choice was due to the Fano resonance spectra exhibited by nano-particles in the toroid area[27], which could affect their intrinsic absorption measurement. By choosing the disk area near the toroidal rim area, we maintained high photothermal sensitivity with minimal differences, allowing accurate measurement of the nanoparticle's absorption based on the photothermal signal. The 2D photothermal image generated through a galvo mirror scan enabled detailed visualization of the absorption properties and specific binding locations of the target nanoparticles on the microtoroid surface. Unlike fluorescence techniques, the photothermal signal arises from the heat dissipation of absorbed light, thereby enabling the detection of non-luminescent materials in the photothermal microscope. Nonetheless, fluorescence images still served as valuable references for photothermal mapping. The detection of single-step quantum blinking behavior from STORM and the intensity distribution of photothermal spots further validated the identification of individual QDs rather than aggregates.

While we have achieved high sensitivity and discrimination capabilities in photothermal microscopy, there are opportunities for improvement. For instance, increasing the phase modulation frequency and adjusting the AM frequency accordingly could reduce the response time and acquisition time of a photothermal image. Moreover, future advancements may involve spectroscopy measurements by varying the pump laser's wavelength or exciting it with different wavelengths to enable multicolor imaging[30]. The integration of photothermal microscopy with WGM resonators opens possibilities for real-time observation of dynamic changes and interactions of target molecules. We believe that overall, FLOWER based photothermal microscopy represents a versatile platform for label-free imaging and single-molecule detection. The demonstrated high sensitivity and discrimination capabilities pave the way for advancements in nanoscale imaging and characterization techniques.

## Methods

**Fabrication of the re-etched microtoroid.** The fabrication process of microtoroid resonators has been previously described[57]. The microtoroid resonators are fabricated on silicon wafers with a 2 µm layer of thermally grown silica. First, circular disc patterns of photoresist with a diameter of 100 µm are created on the top silica layer of the silicon wafer. These photoresist circular pads act as etch masks during immersion in a buffered oxide etchant solution (1:6 V/V) at room temperature, which contains HF (7-15%). After wet etching, any residual photoresist and contaminants are removed using acetone and IPA (isopropyl alcohol). The wafer is then subjected to a post-bake at 130°C to remove moisture[66]. The remaining silica disks act as etch masks during exposure to xenon difluoride (XeF2) gas, resulting in the uniform undercutting of the silica disks and the formation of silicon pillars that support the silica disks. A thermal reflow process using a CO2 laser is employed to shape the silica disk into a microtoroid. Subsequent re-etching of the microtoroid using XeF2 gas is performed to decrease the diameter of the supporting pillars to meet the requirements of the experiment.

**Au nanosphere binds on the microtoroid.** A 100 nm Au nanosphere solution (nanoComposix) was diluted 100 times with HPLC-grade deionized water to achieve a concentration of 5 µg/mL. The diluted Au nanosphere solution was then injected into an aerosol generator. The aerosol generator includes a dryer that removes the liquid water from the Au nanosphere aerosol, resulting in dry aerosol particles. To perform the spraying process, the microtoroid chip was positioned approximately 1 cm below the aerosol output nozzle in a fume hood. This allowed the Au nanospheres to bind on the microtoroid surface.

**Quantum dots functionalization on the microtoroid surface**. After the fabrication of the re-etched microtoroid, the microtoroid chip was cleaned with ethanol and dried by nitrogen gas spray to remove any potential contaminations. The microtoroid chip was then treated with a solution containing 2% v/v of 3-aminopropyl-triethoxysilane (APTES) and ethanol for amine functionalization. The chip was incubated in this solution for 2 minutes at room temperature. After the incubation, the microtoroid chip was rinsed with fresh ethanol and IPA, followed by drying using a flow of nitrogen gas. Next, a mixed QDs solution was prepared with 100 mM1-ethyl-3-(3-dimethylaminopropyl) carbodiimide (EDC) and 100 mM N-Hydroxysulfosuccinimide sodium salt (sulfo-NHS) in 0.1 M 2-(N-morpholino)ethanesulfonic acid (MES) buffer (pH=6.7). The microtoroid chip was then placed in the mixed QDs EDC/NHS solution and incubated for 15 minutes at room temperature. During this incubation, the carboxyl functionalized QDs was bound to the microtoroid surface through the formation of an amide bond. After incubation, the chip was thoroughly rinsed with MES buffer, phosphate-buffered saline (PBS) buffer, deionized water, and ethanol to remove any unreacted reagents or residues. Finally, the chip was dried using nitrogen.

**Fitting.** Spot intensity histograms were analyzed using a Gaussian Mixture Model (GMM)[78,79] which is a parametric probability density function that combines multiple Gaussian component densities with different weights. The histogram was fitted with various numbers of Gaussian components in the GMM using the Expectation-Maximization algorithm[80,81]. To evaluate the fitting performance, the mean value of each peak was examined. It is expected that the mean values of the peaks are the product of the order number and the first peak mean value. After evaluation, it was determined that a GMM with four Gaussian components provided the best fit for the histogram in Fig 4c. For the mixed QDs histogram in Fig 4f, the same methodology is applied, resulting in the identification of five Gaussian distributions.

**STORM fluorescence image.** A N-STORM 5.0 system was used with a CFI HP Apochromat 100X AC TIRF 1.49 NA objective (Nikon) and a 20 mW 405 nm laser unit (LU-NV, Nikon). Following the application of an AT-Qdot 800 filter set (Chroma), the fluorescence signal was captured using a back-illuminated EMCCD (electron-multiplying charge-coupled device) camera (iXon Ultra 897; Andor). The microtoroid chip was imaged in a dry state, positioned upside down on a MatTek dish with a coverslip bottom, and maintained at a room temperature of 22°C. At least 1900 images were acquired to generate the STORM video.

## Acknowledgement
We thank G. Mouneimne for assistance with STORM microscopy.

## Funding
We acknowledge support in part from NIH R35GM137988 and the Gordon and Betty Moore Foundation through Grant GBMF7555.14 to Judith Su.

# Supplementary information to

# Single 5-nm quantum dot detection via microtoroid optical resonator photothermal microscopy

## 1. Configuration of FLOWER based photothermal microscopy

Our photothermal microscopy system is constructed based on an optical sensing system known as FLOWER (frequency locked optical whispering evanescent resonator)[1]. FLOWER tracks the resonance shift of an optical microcavity using frequency locking. Fig. 1a shows a schematic of FLOWER based photothermal microscopy. A tunable external cavity diode laser (New Focus TLB-6712-P Velocity™) is employed as the probe laser. The probe laser output is connected to a lithium niobate phase modulator (PM) (IXblue NIR-MPX800-LN-0.1), which is driven by a 25 MHz oscillation dither signal (DS). A polarization controller (PC) and a 50:50 fiber beam splitter (BS) are utilized to manipulate the phase-modulated probe laser. The PC is responsible for controlling the polarization of the probe laser to maximize the coupling efficiency between the microtoroid and tapered fiber. The BS splits the light into signal and reference arms, which are input separately into a high bandwidth balanced photodetector (New Focus Model 1807). In the signal arm, an optical fiber is thermally tapered using a hydrogen torch to achieve a diameter of approximately 1 μm, ensuring enhanced coupling efficiency[2–4]. The microtoroid optical resonator is evanescently coupled to the system through the tapered signal arm fiber. The output signal from the balanced receiver is multiplied by the dither signal and then time-averaged, resulting in the generation of an error signal as depicted in Fig. 2b. This error signal is proportional to the wavelength detuning between the laser and the microtoroid resonance. To control the laser wavelength, a proportional-integral-derivative (PID) controller is used. The PID controller receives the error signal and provides feedback to the probe laser controller to bring the absolute value of the error signal toward zero. Consequently, the probe laser wavelength is locked precisely at the whispering gallery mode (WGM) resonance wavelength of the microtoroid. By monitoring the output of the PID controller, the resonance wavelength shifts can be accurately measured.

Based on the FLOWER system, a continuous wave (CW) laser (Thorlabs S3FC405) is utilized as the pump laser for particle excitation. To generate a high photothermal signal, a 405 nm wavelength is chosen as the excitation wavelength, as many molecules and particles exhibit characteristic absorptions at the boundary between visible and UV light spectra. The pump laser is subjected to amplitude modulation (AM) using the input signal derived from the function generator. The output of the fiber-coupled pump laser is converted into a free-space beam using a fiber collimator (FC). This collimated pump beam is directed toward a mirror that is mounted on the galvo driver. As the galvo mirror (GM) rotates, the angle of incidence of the collimated beam varies at the back focal plane of the relay lenses. The reflected beam then passes through relay lenses, comprising a scan lens (f = 100 mm) and a tube lens (f = 100 mm). Subsequently, the beam is guided into a 60X objective lens via a cubic beam splitter. The free-space pump laser is focused onto the top surface of the microtoroid. The magnitude of the oscillation in the resonance shift, referred to as the photothermal signal, is directly linked to the heat dissipation resulting from the pump laser. To accurately measure the photothermal signal, a lock-in amplifier (Stanford Research Model SR830) is employed, operating at the frequency of the AM. By utilizing this double lock-in technique, an enhanced signal-to-noise ratio (SNR) of the photothermal image is achieved through a 2D scan of the pump laser.

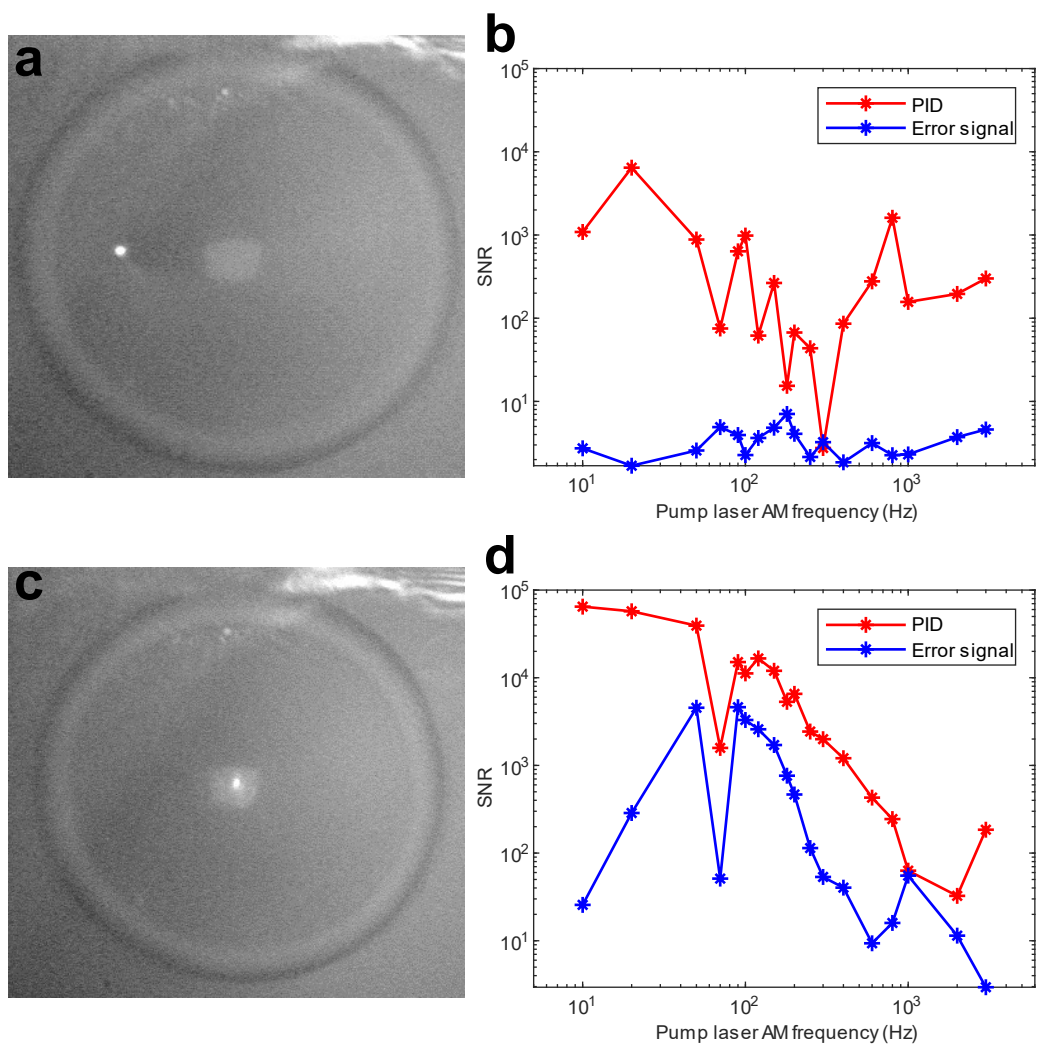

**Figure S1**. Signal-to-noise ratio (SNR) of the photothermal signal. **a**, The pump laser spot is focused on the microtoroid disk area. **b**, Comparison of the SNR between the two photothermal signal sources (either from the PID or from the error signal) for a small photothermal response (at the disk). **c**, The pump laser spot is focused on the microtoroid pillar. **d**, Comparison of the SNR between the two photothermal signal sources for a large photothermal response (at the pillar). In **b** and **d**, PID output signal is input to lock-in amplifier (red curve). The error signal is input to lock-in amplifier (blue curve). The time constant is set at 1 s in the lock-in amplifier.

To mitigate potential damage to the integrated fiber-coupled laser caused by response delays in the constant power control loop, we avoid square wave or TTL modulation applied to the pump laser. Instead, sinusoidal wave AM is employed in the modulation input of the laser. The chosen AM frequency is set at 203.7 Hz. The selection of the modulation frequency requires a careful balance between three objectives: minimizing noise, maximizing the photothermal response of the microresonators, and minimizing the photothermal response time. Low-frequency technical noise, specifically 1/f noise, tends to increase at lower frequencies. By gradually increasing the modulation frequency, the spectral density of 1/f noise can be reduced. However, the re-etched microtoroid exhibits a limited bandwidth of 400 Hz for AM in the photothermal heating signal[5]. It is important to note that the photothermal response time decreases as the AM frequency increases. Taking all these factors into consideration, a reasonable range for choosing the AM signal falls between 200 Hz and 300 Hz. This range strikes a balance among minimizing noise, maximizing photothermal response, and optimizing the photothermal response time.

In our photothermal microscopy configuration, we have two viable options as signal source for the lock-in amplifier: the PID output signal, used as feedback for tuning the probe laser, and the error signal, which is linearly proportional to the detuning of the microtoroid resonance from the probe laser wavelength. Fig. S1 provides a comparison of the SNR for these two photothermal signal sources. The R output of the SR830 lock-in amplifier serves as the photothermal signal, which represents the amplitude of the oscillating resonance shift signal at the AM frequency. In this context, SNR is defined as the ratio of the power of the photothermal signal to the power of the background noise. We note that this is different than from systems whose signal is a power measurement like from a camera or photodiode in which case SNR is $10 \log_{10}$ rather than $20 \log_{10}$.

The microtoroid structure consists of a silica disk supported by a small silicon pillar. When the pump laser scans the bare microtoroid's disk area, the photothermal signal is low due to the low optical absorption of the glass. However, when the pump laser scans the microtoroid pillar, the photothermal signal is high due to the large heat dissipation from the silicon pillar. This leads to a low background signal on the disk and a high background signal on the pillar. Therefore, the effective detection area is the microtoroid disk excluding the pillar.

In Fig. S1b, for small photothermal responses, the SNR of the PID photothermal response is generally higher than that of the error signal photothermal response, except at an AM frequency of 180 Hz. This abnormal decrease in SNR at 180 Hz may be attributed to increased AM noise in the pump laser caused by issues with the laser control circuit. The SNR of the error signal remains below 10 at 180 Hz, resulting in a less noticeable decrease in SNR. A similar abnormal dip in SNR is observed at 70 Hz in Fig. S1d, for both the PID and error signal photothermal responses. In Fig. S1d, for large photothermal responses, the SNR of the PID photothermal response is also higher than that of the error signal photothermal response. Choosing an AM frequency of 203.7 Hz strikes a balance between photothermal SNR and response time. At this frequency, both for large and small photothermal responses, the PID method exhibits a 10 times higher SNR compared to the error signal method. Therefore, the PID photothermal response is more suitable for the re-etched microtoroid.

The re-etched microtoroid, with its small pillar, provides better thermal isolation, resulting in reduced heat dissipation conducted through the pillar to the chip substrate. This improved thermal isolation enhances the sensitivity of the photothermal response. However, it also decreases the microtoroid's AM cutoff bandwidth from 4 or 5 kHz to 400 Hz[5]. When the pump laser operates at low-frequency AM, the probe laser effectively tracks the microtoroid resonance, resulting in a consistently low detuning between their wavelengths. Consequently, the error signal exhibits low sensitivity to the photothermal response. Conversely, the PID output signal is highly sensitive to the photothermal response because it controls the probe laser wavelength. Therefore, for the re-etched microtoroid, the PID photothermal response is a better choice than the error signal method.

## 2. Cross-section absorption of Au nanosphere

The absorption cross section $\sigma_{Au}$ of individual Au nanoparticles is determined based on the Mie theory absorption model[6],

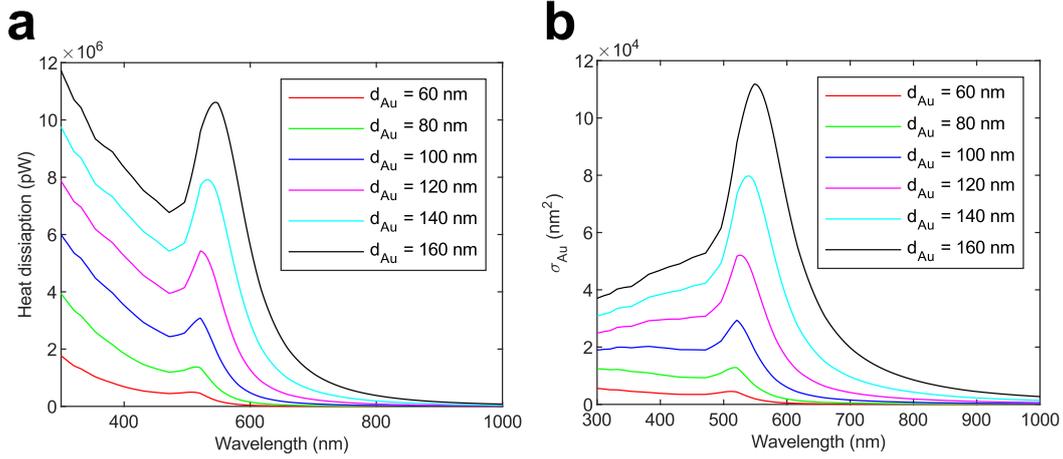

**Figure S2**. Calculated optical and thermal properties of Au nanosphere. **a**, The absorption cross section of Au nanosphere with diameter ranging from 60 nm to 160 nm. At 405 nm, the absorption cross section of 100 nm Au nanosphere is $\sigma_{Au} = 1.9739 \times 10^4 \ nm^2$. **b**, The heat dissipation of the Au nanosphere in **a**. The heat dissipation of the 100 nm Au nanosphere mentioned above under a 405 nm pump laser $P_{heat} = 3.4277 \times 10^{-6} W = 3.4277 \times 10^6 pW$.

$$\sigma_{Au} = \frac{2\pi}{\lambda} Im\{\alpha\}, \tag{1}$$

where $\lambda$ is the wavelength of illuminating light, the corrected polarizability $\alpha$ is[7]

$$\alpha = 3V \frac{1 - (\varepsilon + \epsilon_m)\theta^2/10}{\frac{\varepsilon + 2\epsilon_m}{\varepsilon - \epsilon_m} - \frac{(\varepsilon + 10\epsilon_m)\theta^2}{10} - i\frac{2\epsilon_m^{3/2}\theta^3}{3}} \tag{2}$$

where $\theta = \frac{2\pi r_{Au}}{\lambda}$ is the size parameter and $r_{Au}$ is the Au nanosphere radius. The relative permittivity $\epsilon_m$ is approximately 1 in air. The relative permittivity of Au nanosphere, $\varepsilon$, is taken from Johnson and Christy's report[8].

Fig. S2a shows the absorption cross section $\sigma_{Au}$ of a gold nanosphere. The diameter of the nanospheres ranges from 60 nm to 160 nm. Notably, the absorption cross section exhibits a peak at around 520 nm. As the size of the Au nanospheres increases, the absorption cross section $\sigma_{Au}$ consistently demonstrates an upward trend. In our experiments we used a 100 nm Au nanosphere as our first test particle. The absorption cross section of 100 nm Au nanosphere is $\sigma_{Au} = 1.9739 \times 10^4\ nm^2$ at 405 nm. These absorption cross section results provide insight into the optical properties of the Au nanosphere.

To calculate the heat dissipation of the Au nanosphere, it is essential to determine the intensity of the pump laser spot. The diameter of the pump laser can be calculated using the Rayleigh criterion, which states that:

$$d_{spot} = 1.22 \times \frac{\lambda}{NA}, \tag{3}$$

where the pump laser wavelength $\lambda = 405\ nm$, and $NA$ is the numerical aperture of the pump light illumination system. The spot intensity is:

$$I_0 = \frac{P_0}{S_{spot}} = \frac{4P_0}{\pi d_{spot}^2}, \tag{4}$$

where $P_0$ is the power of the laser spot focused on the microtoroid top surface. The 405 nm pump laser spot intensity is $17.4\ KW/cm^2$. The target particle absorbed power equals to the heat dissipation of the Au nanosphere,

$$P_{heat} = P_{abs} = I_0\ \sigma_{au} \tag{5}$$

The heat dissipation of the Au nanosphere is calculated based on its absorption cross section, as shown in Fig. S2a, using Eq. (5). The heat dissipation of the Au nanosphere under 405 nm pump laser excitation is illustrated in Fig. S2b. Specifically, for a 100 nm Au nanosphere, the heat dissipation is calculated to be $P_{heat} = 3.4277 \times 10^{-6} W = 3.4277 \times 10^6 pW$ under a 405 nm pump laser. It is worth noting that the microtoroid has the capability to detect heat dissipation power of a few tens of pW[9]. Therefore, the 100 nm Au nanosphere exhibits a significant photothermal response making it easily detectable.

## 3. Characterizations of QDs

When considering luminescent particles as the target, the absorption calculation presents a different scenario, as a portion of the absorbed light energy is transferred to emitted light. In this case, the fraction of heat dissipation in fluorescence particles is:

$$\eta_{heat} = 1 - \eta_{fl} + \frac{\eta_{fl}\left(\frac{1}{\lambda_{exc}} - \frac{1}{\lambda_{fl}}\right)}{\frac{1}{\lambda_{exc}}} = 1 - \eta_{fl}\frac{\lambda_{exc}}{\lambda_{fl}}, \tag{6}$$

where $\eta_{fl}$ is the quantum yield of the fluorescence particle, $\lambda_{exc}$ is the excitation wavelength, and $\lambda_{fl}$ is the emission wavelength. The absorbance of the fluorescence particle solution was measured using a Nanodrop. According to the Beer-Lambert law, the molar extinction coefficient $\varepsilon_{ext}$ is calculated in units $M^{-1}cm^{-1}$,

$$\varepsilon_{ext} = \frac{A}{cl}, \tag{7}$$

where $A$ is the absorbance of the fluorescence particle solution, $c$ is the molar concentration of the absorbing particles, $l$ is the path length of the sample solution in unit $cm$. The cross-section absorption $\sigma_{abs}$ is in unit $cm^2$, and can be calculated by,

$$\sigma_{abs} = \varepsilon_{ext} \times 10^{-1} \times In(10)/N_A \tag{8}$$

in unit of $nm^2$, where the Avogadro constant, $N_A = 6.02 \times 10^{23}$. Finally, the heat dissipation is,

$$P_{heat} = P_{abs} \times \eta_{heat} \tag{9}$$

The Qdot 800 QDs, supplied by Thermo Fisher Scientific have a diameter ranging from 18 nm to 20 nm. According to the certificate of analysis, these QDs have a quantum yield of 62% and emit light at a wavelength of 793 nm. On the other hand, the quantum yield of DiagNano 800 QDs, supplied by CD Bioparticles, was not provided, so it was determined by comparing the emission intensity of DiagNano 800 with that of Qdot 800,

$$\eta_2 = \frac{I_2}{\varepsilon_2} \times \frac{\varepsilon_1}{I_1} \times \eta_1, \tag{10}$$

where $\eta$ is the quantum yield, $\varepsilon$ is the molar extinction, $I$ is the fluorescence emission intensity, with subscripts 1 and 2 representing DiagNano 800 and Qdot 800, respectively. The fluorescence emission intensities of the QDs were measured in the fluorescence image. The quantum yield of DiagNano 800 QDs was calculated to be 15.64%. These nanoparticles have a diameter ranging from 5 nm to 6 nm and exhibit similar fluorescence properties to Qdot 800, emitting light at a wavelength of 800 nm. The characteristic information of the two QDs used in the photothermal map is presented in Table 1. The heat dissipation of Qdot 800 is 363.1 pW, while the heat dissipation of DiagNano 800 is 71.25 pW. The ratio of their heat dissipation is approximately 5.10. In the experiment, the photothermal response ratio between the two QDs is approximately 5.36. It is important to note that the photothermal response is proportional to the heat dissipation. Therefore, the photothermal response ratio should remain consistent with the heat dissipation ratio.

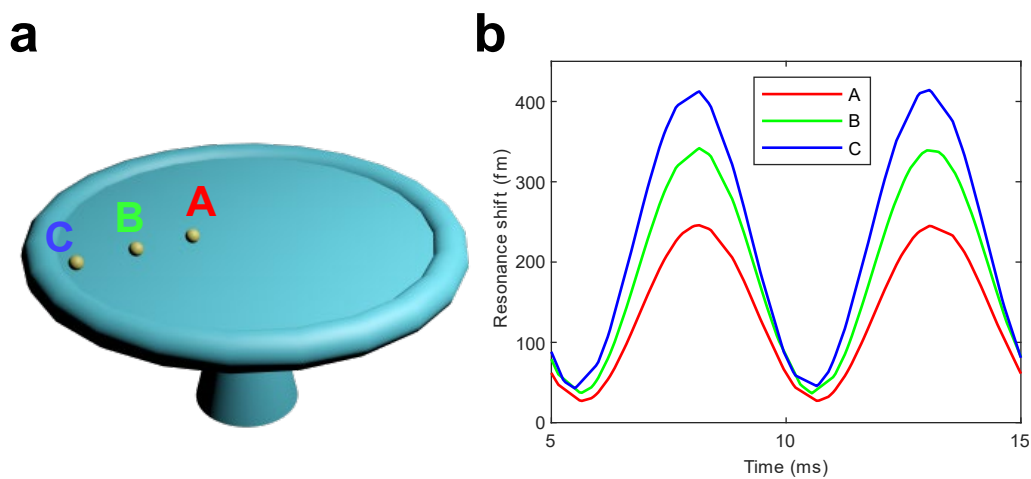

**Figure S3** COMSOL simulation of the microtoroid resonance shift. **a**, The rendered model of the microtoroid with 100 nm Au nanospheres positioned on the microtoroid resonator. The microtoroid resonator has a major diameter of 70 µm and a minor diameter of 3 µm. The Au nanospheres A, B, and C are individually placed at distances of 10 µm, 20 µm, and 30 µm from the center of the microtoroid disk. **b**, The observed resonance shift resulting from the absorption of the 405 nm pump laser by the Au nanospheres A, B, and C in **a**.

## 4. WGM resonance shift COMSOL simulation

The photothermal response is characterized by an oscillating resonance shift signal resulting from the absorption of the AM pump laser by the target particle. In COMSOL simulation, the 100 nm Au nanoparticles serve as the heating source, with previously calculated heat dissipation of $3.4277 \times 10^6 \; pW$ using the experimental 405 nm pump laser. The pump laser spot size is larger than the size of the target particles. As a result, a significant portion of the light illuminates the top surface of the microtoroid. Since silica material exhibits minimal absorption, the light absorbed by the silica microtoroid can be neglected compared to the absorbed light by the particles. Most of the pump light is absorbed by the nontransparent silicon chip substrate, which is significantly larger in size compared to the microtoroid (with a major diameter of approximately 70 µm). Most of the heat dissipation in the substrate readily conducts throughout the entire chip substrate. However, due to the small diameter (below 5 µm) of the microtoroid's silicon pillar, the contribution of heat dissipation from the chip substrate to the photothermal signal is minimal. The simulation employs the finite-element analysis method. In the simulation, the heat dissipation process of the Au nanoparticles is simplified as point heat sources. The resonance wavelength shift signal is determined by the resonance condition of the optical resonator, which is defined by the resonance condition of optical resonator,

$$\lambda_{res} = \frac{OPL}{m} \tag{11}$$

where $\lambda_{res}$ is the resonance wavelength, $OPL$ is the optical path length (OPL), $m$ is the mode number of WGM. The OPL is calculated by integrating the product of the geometric length of the optical path followed by the WGM and the refractive index of the medium. The time-dependent simulation considers both heat transfer and thermal expansion simultaneously. The microtoroid's deformation caused by the thermal expansion from the point heat source is utilized for the calculation of the geometric length. The thermo-optic coefficient ($dn/dT$) of fused silica at room temperature (295 K) is $8.6 \times 10^{-6}\ K^{-1}$. The refractive index information is acquired by the temperature distribution on the microtoroid rim. The resulting resonance shift of the 100 nm Au nanosphere at different positions on the microtoroid is illustrated in Fig. S3b. In the microtoroid's disk area, the resonance shift is larger when the particle is positioned closer to the microtoroid's rim. The photothermal sensitivity of the nanoparticle exhibits a limited difference between positions B and C within the disk area. This limited area with a well-defined photothermal spot is suitable for measuring the particle's absorption cross section through the photothermal response.

## 5. Photothermal spot of single QDs

For the analysis of photothermal spot intensities of individual quantum dots, the Gaussian Mixture Model (GMM) is employed photothermal spot peak distribution. This enables the assessment of the photothermal intensity distribution of the single quantum dots. Subsequently, a specific quantum dot is located and selected for a high-resolution scan (x direction: 75 nm/pixel, y direction: 9.375 nm/pixel). The dissimilarity in pixel resolution between the x and y directions arises from the microtoroid's 2D scanning procedure, which is carried out line by line along the y direction. Although the data acquisition card efficiently acquires data at a high sample rate for the y direction, the x direction is constrained by the number of scan lines. Enhancing resolution in the x direction would substantially extend the scan time. Consequently, during the scan, the pixel resolution in the y direction is design to be higher than that in the x direction due to the sample rate of the lock-in amplifier output signal.

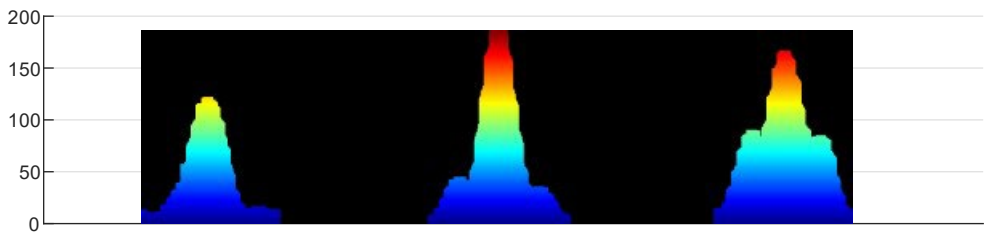

**Figure S4 Photothermal mapping of a single 5-6 nm QD at different focus positions**.

**a**, Photothermal images of the same DN 800 QD (size: 5-6 nm) with the background signal removed. 1, The QD is positioned at d = -2 µm, out of focus and behind the focal point of the objective lens. 2, The QD is d = 0 precisely at the focus point. 3, The QD is located at d = 1 µm, appearing blurred in front of the focal point. Scale bar, 1 µm. **b**, Photothermal signal profiles and scans of the same DN 800 QD at varying distances from the focal point, corresponding to the photothermal maps 1-3 in **a**.

The results of photothermal mapping of single QDs at different focus positions are depicted in Fig. S4a. This photothermal map undergoes background signal subtraction, effectively yielding a final photothermal map with the original photothermal signal and estimated background removed. Such subtraction enhances the visibility of significant features within the data by minimizing the influence of background noise Firstly, we create a disk-shaped structuring element, often referred to as a kernel or mask, to define a local neighborhood around each pixel in the dataset. Next, we employ the "imopen" function to execute morphological opening on this disk-shaped structuring element. Morphological opening, a mathematical morphology operation, comprises two steps: erosion followed by dilation. During erosion, the structuring element is centered over each pixel in the input data, leading to the reduction of small bright spots and an overall decrease in image intensity. Then, dilation is applied with the structuring element centered over each pixel. This step enlarges the remaining features, emphasizing large structures. As a result of these two data processing steps, we acquire the background signal. Finally, by subtracting this background dataset, photothermal maps without the background are shown in Fig. S4a. This enables us to focus on the specific feature spots of interest, namely the photothermal spots of single QDs.

To directly illustrate the influence of focus conditions on photothermal spot intensity, Fig. S4b displays the profiles of the spots 1-3 with varying focus positions from Fig. S4a. In the initial scan, it's common for the quantum dot to be off the focus plane. As seen in Fig. S4a, the photothermal map of the first scan is depicted in Fig. S4a1, where the quantum dot is positioned out of focus and behind the focal point. Here, the photothermal peak value is measured at 117.76 µV. With subsequent scans and focal plane adjustments, the photothermal map at the focus plane is presented in Fig. S4a2, showing photothermal peaks at 186.26 µV. In contrast, Fig. S4a3 portrays the quantum dot at a distance of d = 1 µm, resulting in a blurred appearance in front of the focal point. Consequently, the photothermal peak value diminishes to 162.19 µV. It's important to comprehend that the photothermal signal represents the amplitude of the oscillating resonance shift signal. The SNR is defined as the ratio of the power of the photothermal spot's peak to the power of the encompassing background noise. Under optimal focus conditions, the SNR of the photothermal signal emanating from the 5 nm DN 800 QD culminates at $1.17 \times 10^4$.

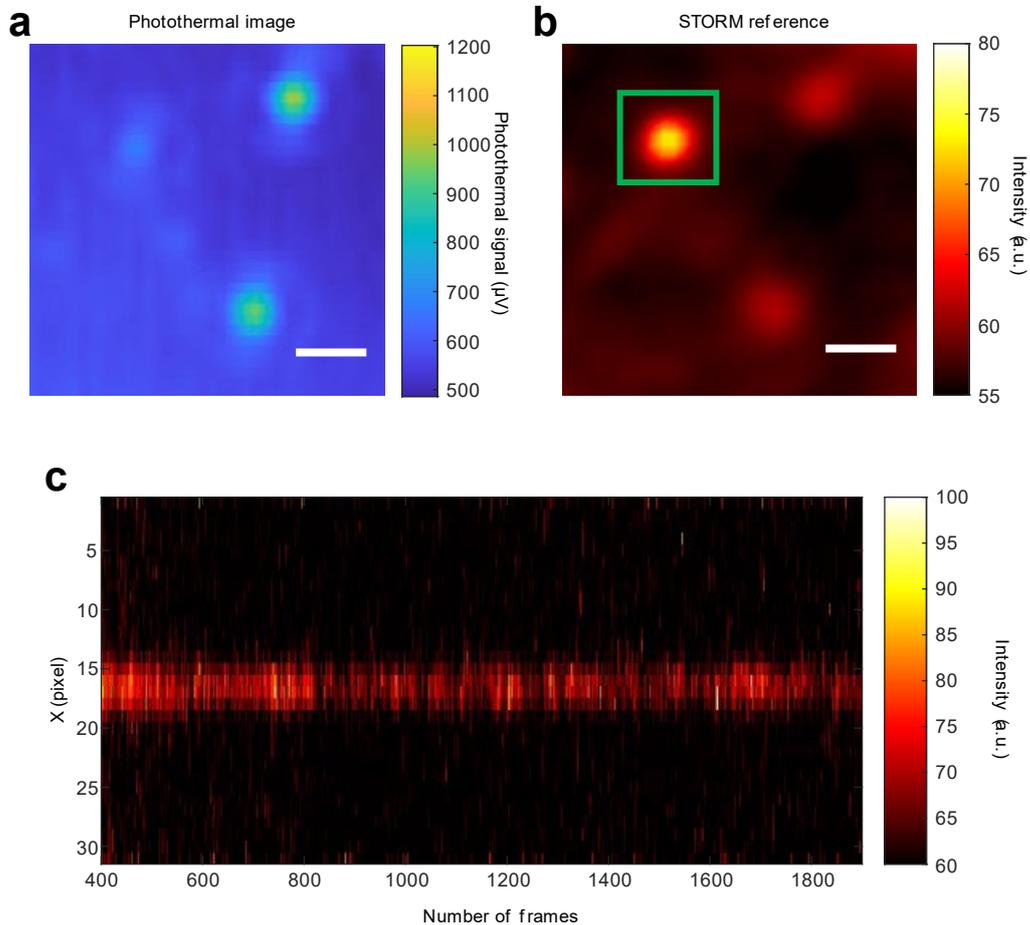

Figure S5 Photothermal image of single QDs. **a**, Photothermal scan of Qdot 800 QDs (size:18~20nm). **b**, Corresponding STORM fluorescence reference image of **a**. **c**, The photo blinking behavior of single QDs within the marked green square in **b**.

## 6. STORM fluorescence image QDs

To validate the accuracy of the photothermal images of QDs, we used fluorescence images of the QDs as references for the photothermal map. In our experimental procedure, after functionalizing the QDs onto the microtoroid, we executed a photothermal scan of the microtoroid, yielding the photothermal image of Qdot 800 QDs as depicted in Fig. S5a. Subsequently, the microtoroids containing the QDs were imaged using Stochastic Optical Reconstruction Microscopy (STORM). This STORM imaging involved a CFI HP Apochromat 100X AC TIRF 1.49 NA objective with a 405 nm laser for excitation. The Qdot 800 QDs (size: 18~20 nm) were successfully observed in the STORM image which is shown in Fig. S5b. Both images reveal identical positions of the three QDs spots. A one-minute video is recorded in the STORM. The single step photon blinking was observed in the STORM video and demonstrated in Fig. S5c. However, due to the lower quantum yield and smaller size, the single DN 800 QDs (size: 5-6 nm) were not observable through STORM. In the case of DN 800 QDs bound to the microtoroid, the count of fluorescence spots was much lower compared to the spots in photothermal image. This difference can be attributed to the fact that the single DN 800 QDs can't emit sufficient light for detection through STORM. Although, there may exist some single DN 800 QDs that are observable via STORM with an extended exposure time, their emitted signals remain too weak to observe the blinking effect in a suitable exposure time.